\documentclass[11pt,twoside]{article}  

\usepackage[utf8]{inputenc}
\DeclareUnicodeCharacter{2217}{*}
\usepackage{amsfonts}
\usepackage{physics}
\usepackage{multicol}
\usepackage{amssymb}
\usepackage{amsmath}
\usepackage{amsthm}
\usepackage[english]{babel}   
\usepackage{setspace}
\usepackage{cite}
\usepackage{subcaption}

\title{Bounding light source side channels in QKD via Hong-Ou-Mandel interference}

\author{A. Duplinskiy$^{1,2,3}$, D. Sych$^{3,4,5}$}

\date{\small
$^1$ Moscow Institute of Physics and Technology, 9 Institutskiy per., Dolgoprudny, Moscow Region, 141701, Russian Federation\\%
$^2$ Russian Quantum Center, Skolkovo, Moscow 143025, Russia\\%
$^3$ QRate, Skolkovo, Moscow 143025, Russia\\%
$^4$ P. N. Lebedev Physical Institute, Russian Academy of Sciences, Leninskiy Prospekt 53,
Moscow, 119991, Russia \\%
$^5$  Department of Physics, Moscow Pedagogical State University, 29 Malaya Pirogovskaya St, Moscow, 119435, Russia \\[2ex]%
duplinskii@phystech.edu\\[2ex]%
}

\usepackage[pdftex]{graphicx}
\usepackage{multicol}
\usepackage[left=2cm,right=2cm,
top=2cm,bottom=2cm,bindingoffset=0cm]{geometry}
\begin{document}
\maketitle

Side-channel attacks on practical quantum key distribution systems compromise its security. Although some of these attacks can be taken into account, the general recipe of how to eliminate all side-channel flaws is still missing. In this work, we propose a method for estimation of the total passive side-channel information leakage from the Alice's light source. The method relies on Hong-Ou-Mandel interference between different signals emitted by Alice, which reveals their overall mode mismatch without the necessity to measure all individual degrees of freedom independently. We include experimental values of interference visibility in the security proof for the decoy-state BB84 protocol, and lower-bound the secure key rate for realistic light sources. The obtained results provide a tool that can be used for certification of the current QKD systems and pave the way towards the loophole-free design of the future ones.

\section*{Introduction}
Quantum key distribution (QKD) allows unconditional secrecy of sharing  keys  between  distant  users  –  the  transmitter  (Alice)  and  the  receiver  (Bob). Although QKD is guaranteed to be theoretically secure, there is a significant mismatch between the theoretical models and the real-world QKD systems. Even though many of the device imperfections have been successfully included in the security proofs, drawbacks still exist and need to be covered. 

One of the issues that is not yet fully taken into account is side-channel information leakage from the pulses emitted by Alice. Normally, QKD protocols assume that signals are completely indistinguishable beyond the operational space which is used for encoding secure bits. This assumption, however, does not usually hold in practice. Even if two lasers seem identical, actual parameters of the emitted light can vary. Temporal, spatial and spectral differences of the pulses from independent sources have been experimentally studied in the work \cite{nauerth2009}. Authors quantified mutual information between side-channel parameters and bit values without addressing security proofs. Later on, based on Koashi security proof \cite{koashi2009}, a model of side channels for the Trojan-horse attack was developed \cite{lucamarini2015, tamaki2016}. This model can be adopted for passive side-channel effects estimation.

In this regard, the most vulnerable type of setup is a multiple laser source system.  For example, polarization encoding is widely used for free space QKD, as atmosphere induces only minor disturbance of the polarization state of light. The simplest way to implement polarization encoding for BB84 protocol at gigahertz frequencies  is to use four different lasers, one for each state \cite{hughes2002,schmitt2007,garcia2013,vallone2014}. Furthermore, there are many fiber-channel setups that are based on multiple lasers, some of them implement quantum cryptography beyond QKD \cite{peng2007,yin2017sign}.

Previous work propose individual measurements of different degrees of freedom to access information about pulses' similarity. To prove unconditional security, however, one needs to measure all possible characteristics, whereas individual measurements of the entire set of separate parameters can be highly complicated. Moreover, the identical partial distributions do not guarantee that the joint distributions are also identical.
For example, the same temporal and spatial distributions do not guarantee that the joint distributions of intensity in time and space also match. As well as spectral distribution may depend on time due to the frequency chirp of a laser source \cite{koch1984}.

In contrast to individual measurements of separate degrees of freedom, the effect of interference allows to study the overall distinguishability of the pulses, since distinguishable states do not interfere \cite{feynman1965}. For example, fourth-order, or Hong-Ou-Mandel (HOM), interference is a standard method for characterization of single-photon sources \cite{hong1987,osorio2013}. The possibility of utilizing HOM interference to characterize unmeasured degrees of freedom for the QKD sources with multiple lasers has been mentioned in work \cite{yuan2014}. 

In this work, we propose to use HOM interference of phase-randomized weak coherent pulses (PRWCPs) to study their distinguishability, and upper-bound the influence of side-channel effects that can possibly compromise the security of QKD. We calculate the key generation rate depending on the visibility between different states and show the practical applicability of the proposed method.

The paper is organized as follows. In Sec.~\ref{HOM_part} we describe Hong-Ou-Mandel effect for PRWCPs from the quantum point of view, and link visibility of HOM interference with distinguishability of interfering states. Next, in Sec.~\ref{basis_flaw_part} the results of the HOM interference are used to calculate the value of bases imbalance. Then, in Sec.~\ref{keyrate_part} the value of bases imbalance are used to correct estimation of single-photon error rate, and, finally, obtain key generation rate for different values of HOM visibility. Based on  currently known experimental results, applicability of the proposed method is discussed in Sec.~\ref{discussion_part}.

\section{Hong-Ou-Mandel interference and non-orthogonality of quantum states} \label{HOM_part}

\begin{figure} [h]
	\centering
	\includegraphics[width=0.8\textwidth]{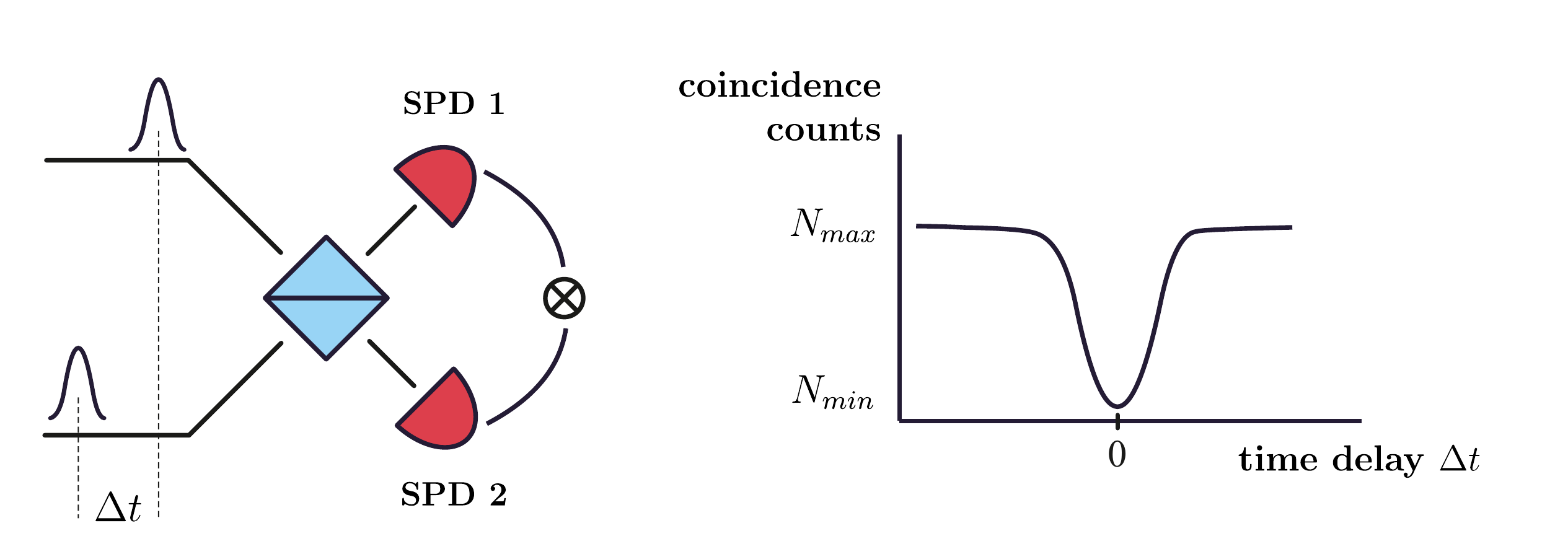}
	\caption{Typical setup for HOM experiment (on the left). Two single-photon pulses enter a beamsplitter, while coincidence counts are measured with two single-photon detectors (SPDs). Dependence of the number of coincidence counts on the time delay $\Delta t$ has a dip that corresponds to the maximum overlap of photons on the beamsplitter (on the right).}
	\label{HOM_classic}
\end{figure}

The fourth-order interference of single photons has been introduced by Hong, Ou and Mandel in 1987 in order to measure the time delay between single photons precisely \cite{hong1987}. The key idea is that two indistinguishable single photons matched on a 50:50 beam-splitter always exit it pairwise. Experimental setup to observe the effect is illustrated in the Fig. \ref{HOM_classic}. 

Single photon detectors in both arms allow to count coincidence clicks, when two photons leave the beamsplitter in different arms. While the time difference $\Delta t$ between single-photon wavepackets is large enough so that the photons do not overlap, their behaviour is independent from each other. As a result, coincidence clicks are observed in half of the detection events. 

As soon as overlap becomes non-zero, the number of coincidence click decreases. In the limit of perfectly mode-matched pure single-photon states, coincidence clicks completely vanish. In order to characterise the indistinguishability of photons, the visibility of HOM-interference is introduced as the difference between the maximum and minimum numbers of coincidence clicks, divided by the maximum number of coincidence clicks: $V = (N_{max}-N_{min})/N_{max}$ (Fig. \ref{HOM_classic}). It is easy to see, that visibility equals 0 for completely orthogonal (distinguishable) states and equals 1 in case of perfect indistinguishability. In case one knows the density matrices of the single-photon states ($\hat{\rho}_1^{\text{sp}}$ and $\hat{\rho}_2^{\text{sp}}$), the visibility can be calculated as:
\begin{equation} \label{single_ph_vis}
\operatorname{V}_{\hat{\rho}_1^{\text{sp}} \hat{\rho}_2^{\text{sp}}}=\Tr\left(\hat{\rho}_1^{\text{sp}} \hat{\rho}_2^{\text{sp}}\right).
\end{equation}

Most of the practical QKD systems utilize phase randomised weak coherent pulses (PRWCPs) instead of the single-photon states. 
Nevertheless, a HOM-like effect also takes place for two PRWCPs as well, though the coincidence clicks never totally disappear. The maximum possible visibility obtained in this type of experiment is 0.5. In contrast to the single-photon case, the HOM effect for PRWCPs can be explained both from classical and quantum points of view. It is quite natural as coherent states could be treated as the ``most classical'' quantum states. Anyway, in this paper we focus on the quantum approach and study how the visibility depends on the characteristics of interfering states.

The PRWCPs can be expressed as a Poissonian combination of independent Fock states \cite{allevi2013}. Let us consider two such states $\hat{\rho}_1$ and $\hat{\rho}_2$ with equal intensity $\mu$ at a 50:50 beamsplitter. Their density matrices can be written in Fock basis as follows:

\begin{equation} \label{rho}
\hat{\rho}=\sum_{n=0}^{\infty} \rho_{nn} \ket{n} \bra{n} = \sum_{n=0}^{\infty} e^{-\mu^2} \frac{\mu^n}{n!} (\hat{a}^{\dagger})^n\ketbra{0}{0}(\hat{a})^n.
\end{equation}



To simulate a real system one has to take into account detection efficiency, including both optical losses and single-photon detector efficiency. Here we consider $\mu$ as effective intensity of states, corrected for detection imperfections. Thereby the input state on a beamsplitter is:
\begin{multline} \label{rho_in}
\hat{\rho}_{in} = \hat{\rho_1} \otimes \hat{\rho_2}= e^{-\mu^2} \Biggl(\ketbra {0_1 0_2}{0_1 0_2} + \mu \biggl( \ketbra {0_1 1_2}{1_1 0_2} + \ketbra {1_1 0_2}{0_1 1_2} \biggr) + \\
+\frac{\mu^2}{2}\biggl( \ketbra{0_1 2_2}{0_1 2_2}+2\ketbra{1_1 1_2}{1_1 1_2}+\ketbra{2_1 0_2}{2_1 0_2}\biggr)+O(\mu^3)\Biggr).
\end{multline}
We omit the terms higher than the second degree, as we consider $\mu$ to be small. As soon as the terms with the total photon number less than two do not contribute to coincidence clicks, the only states that we need to study are:
\begin{equation}
\hat{\rho}_{in}^{'}= \frac{e^{-\mu^2} \mu^2}{2}\biggl( \ketbra{0_1 2_2}{0_1 2_2}+2\ketbra{1_1 1_2}{1_1 1_2}+\ketbra{2_1 0_2}{2_1 0_2} \biggr).
\end{equation}
In case of orthogonal modes, all three terms: $\ketbra{0_1 2_2}{0_1 2_2}$, $\ketbra{1_1 1_2}{1_1 1_2}$ and $\ketbra{2_1 0_2}{2_1 0_2}$ contribute equally to coincidence clicks. 
The middle term $\ketbra{1_1 1_2}{1_1 1_2}$ leads to the standard single-photon HOM effect, i.e. for perfect node-matching coincidences vanish. At the same time, coincidence clicks due to the other terms: $\ketbra{0_1 2_2}{0_1 2_2}$ and $\ketbra{2_1 0_2}{2_1 0_2}$ never change, since there is no interference with vacuum.
Thus the maximum visibility value is 0.5. In the intermediate case of partial distinguishability, the number of coincidence clicks from the state $\ketbra{1_1 1_2}{1_1 1_2}$ reduces by the factor of $(1 - \operatorname{V}_{\hat{\rho}_1^{\text{sp}} \hat{\rho}_2^{\text{sp}}})$, according to the (\ref{single_ph_vis}), resulting in visibility for PRWCPs

\begin{equation} \label{visibility}
\operatorname{V}_{\hat{\rho}_1 \hat{\rho}_2} = \frac{1}{2} \operatorname{V}_{\hat{\rho}_1^{\text{sp}} \hat{\rho}_2^{\text{sp}}}= \frac{1}{2} \Tr\left(\hat{\rho}_1^{\text{sp}} \hat{\rho}_2^{\text{sp}}\right).
\end{equation}

\begin{figure} [h]
\centering
\includegraphics[width=0.9\textwidth]{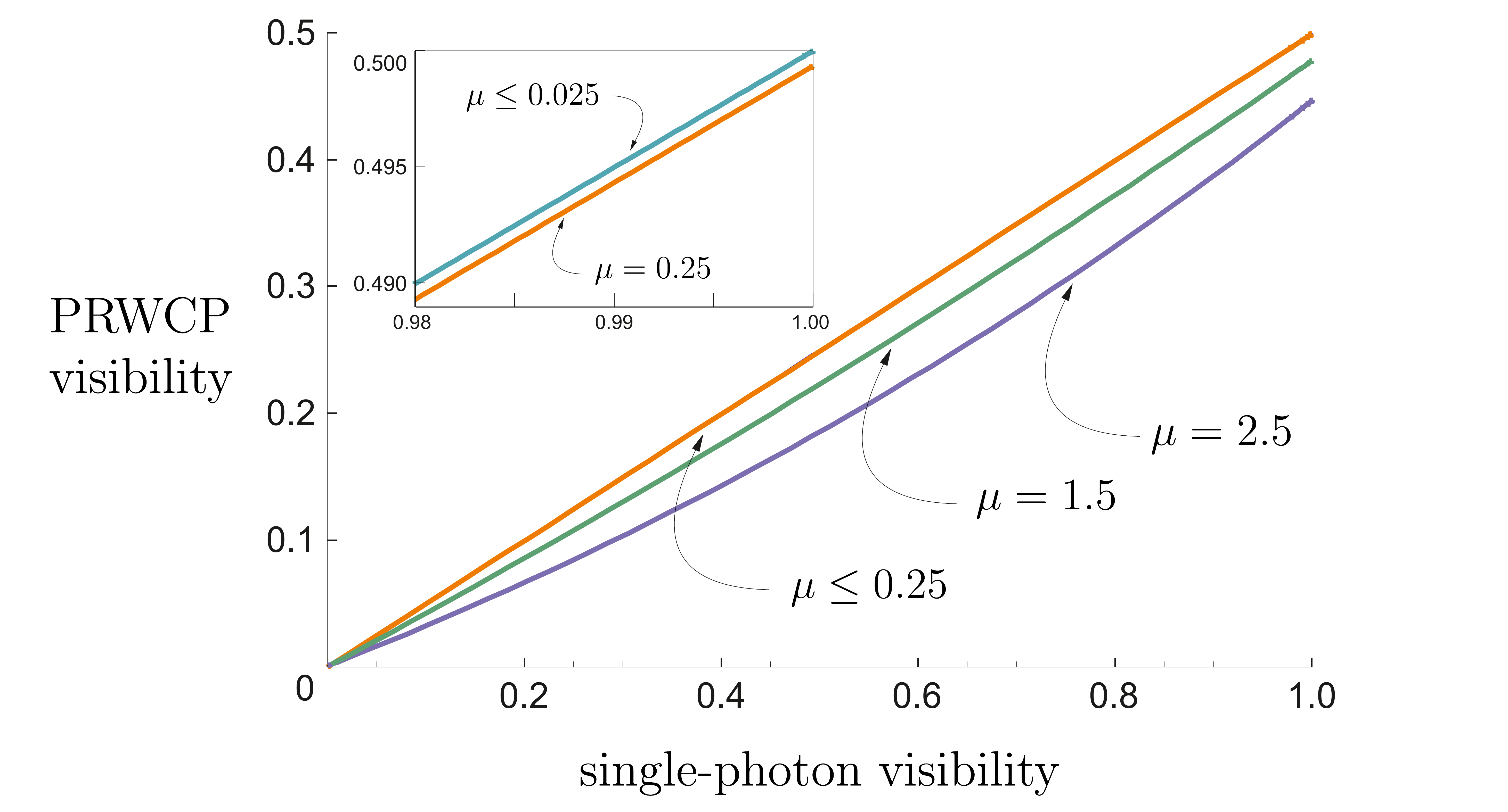}
\caption{Simulation results for HOM visibility of phase randomized weak coherent pulses (PRWCPs), depending on the HOM visibility for single-photon states within the PRWCPs. Plot shows the dependence for different mean photon numbers per pulse $\mu$. Simulation is done including the terms (\ref{rho_in}) up to the 20-th degree. }
\label{HOM}
\end{figure}

In these calculations we neglect high order terms. To check the applicability limit of the obtained result, we carry out numerical simulations, including terms up to the 20-th degree. Simulation results (Fig. \ref{HOM}) show that for $\mu$ below 0.025 photons per pulse the equation (\ref{visibility}) is very close to the actual behaviour. Even for higher values of $\mu$ (e.g. 0.25) the dependence remains very close to linear, however the visibility limit should be slightly corrected. For $\mu$ value more than 1 photon per pulse Eq. \ref{visibility} is no longer applicable.

\section{Side-channel information and HOM visibility} \label{basis_flaw_part}

Side channel is a collective name for vulnerabilities that can cause information leakage to Eve, bypassing the communication protocol. Here we address only passive type of side-channels for Alice's device, meaning that information can be partially revealed via the distinguishability of pulses in non-operational degrees of freedom. In other words, we consider mode mismatch between different optical signals, emitted by Alice.


We address conventional BB84 protocol with decoy states and study distinguishability between the two bases \cite{koashi2009,lucamarini2015}, caused by the non-operational degrees of freedom. This approach is sufficient to quantify all BB84 vulnerabilities caused by distinguishability, as it has been proven, that for identical bases density matrices all bit flaws cause an increase of the respective error rate \cite{koashi2003}.



Ideally, the protocol implies that Eve is not able to discriminate one basis from the other, as their density matrices  $\hat{\rho}_x = \frac{1}{2} (\ketbra{H} + \ketbra{V})$ and $\hat{\rho}_z = \frac{1}{2} (\ketbra{D} + \ketbra{A})$ are supposed to be the same. Non-perfect mode matching between the different bits causes the differences in the bases density matrices, resulting in a vulnerability that could be exploited by Eve. The value that allows to quantify the impact of this effect is so-called bases imbalance \cite{koashi2009,lucamarini2015}:

\begin{equation}
\Delta = \frac{1- \sqrt{\operatorname{F}(\hat{\rho}_x, \hat{\rho}_z)}}{2},
\end{equation}
Where $\operatorname{F}(\hat{\rho}_x, \hat{\rho}_z)$ is fidelity between the density matrices \cite{uhlmann1976, jozsa1994}
\begin{equation} \label{fidelity}
\operatorname{F}(\hat{\rho}_x, \hat{\rho}_z)=\left(\operatorname{Tr} \sqrt{\sqrt{\hat{\rho}_x} \hat{\rho}_z \sqrt{\hat{\rho}_x}}\right)^2 = \Big(\operatorname{Tr}|\sqrt{\hat{\rho}_x}\sqrt{\hat{\rho}_z}|\Big)^2.
\end{equation}
It is easy to see that in the perfect case, when $\hat{\rho}_x$ and $\hat{\rho}_z$ are indistinguishable, $\Delta$ equals 0, whereas in the worst case scenario,  it reaches its maximum value 1/2.

We can express density matrices of X and Z bases as sums of bits, each of which is a tensor product of density matrices in operational and non-operational degrees of freedom:

\begin{equation}
\hat{\rho}_x = 
\frac{1}{2}\mqty(\dmat{1 & 0 \\ 0 & 0}) \otimes \hat{\rho}_{x, 0}^{\lambda} +
\frac{1}{2}\mqty(\dmat{0 & 0 \\ 0 & 1}) \otimes \hat{\rho}_{x, 1}^{\lambda},
\end{equation}

\begin{equation}
\hat{\rho}_z = 
\frac{1}{2}\mqty(\dmat{ 1/2 &  1/2 \\ 1/2 &  1/2}) \otimes \hat{\rho}_{z, 0}^{\lambda} +
\frac{1}{2}\mqty(\dmat{ 1/2 &  - 1/2 \\ - 1/2 &  1/2}) \otimes \hat{\rho}_{z, 1}^{\lambda},
\end{equation}
where $\hat{\rho}_{b, i}^{\lambda} $ accounts for non-operational degrees of freedom for the bit $i$ in the basis $b$. 

\begin{figure}[h]
	\centering
	\includegraphics[scale=0.4]{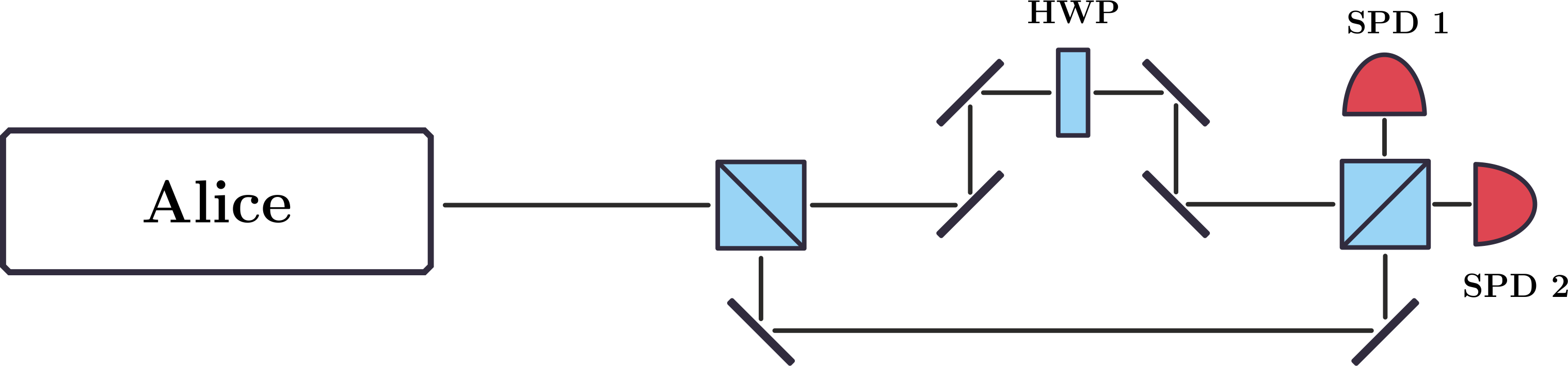}
	\caption{\small Concept of a possible optical scheme, that can be used for HOM-based basis test of polarization-encoding system. One of the interferometer arms  is delayed, so that two consequent pulses are matched on a beam-splitter. Half-wave plate is used for polarization matching of two states.}
	\label{basis_test}
\end{figure}

In Appendix B we construct a lower bound on fidelity between X and Z bases depending on the fidelity between density matrices of non-operational degrees of freedom. Then, using the triangle inequality for Bures angle \cite{ma2009fidelity}, we obtain the result: 

\begin{equation} \label{finalFid}
\arccos \sqrt{\operatorname{F}(\hat{\rho}_x, \hat{\rho}_z)} \leq \operatorname{arccos}\max_{\scriptscriptstyle i,j \in {0, 1}} \sqrt{F(\hat{\rho}_{x, i}^{\lambda}, \hat{\rho}_{z, j}^{\lambda})} + \sum_{ \scriptscriptstyle b \in \left\lbrace x, z \right\rbrace  }  \operatorname{arccos}\frac{1+\sqrt{F(\hat{\rho}_{b, 0}^{\lambda}, \hat{\rho}_{b, 1}^{\lambda})}}{2}.
\end{equation}
Here the first term indicates the difference between the bases, whereas the second term represents differences between the bits within bases. This result allows us to estimate bases imbalance with only pairwise interference experiments with separate pulses. To conduct an interference experiment on non-operational degrees of freedom, we only need to match the bits in operational space. Fig. \ref{basis_test} illustrates an example a of possible optical setup that could be used to carry out such kind of experiment for a QKD system with polarization encoding. A delayed Mach-Zender interferometer allows two consequent pulses to overlap on a beamsplitter with two single-photon detectors (SPD) at the output. A half-wave plate is used to match the polarization of different bits. It can also be used instead of the time delay to vary the degree of mode overlap.

Finally, we establish a connection between visibility and fidelity. In Appendix A we derive an equation to calculate fidelity for two arbitrary PRWCPs ($\hat{\rho}_1, \hat{\rho}_2$) with equal mean photon number $\mu$  depending on their HOM visibility: 
\begin{equation}
\sqrt{F(\hat{\rho}_1, \hat{\rho}_2)} = \exp(\mu(\sqrt{2\operatorname{V}_{\hat{\rho}_1 \hat{\rho}_2}}-1)).
\end{equation}

We substitute this result into equation (\ref{finalFid}). In principle, every two pulses have their own value of visibility, but to simplify the visualization of the final result, we demonstrate the special case when every pair of bits has the same visibility value V:

\begin{equation} 
1 -2\Delta \geq   \operatorname{cos} \Bigg(2 \operatorname{arccos}\frac{1+e^{\mu(\sqrt{2V}-1)}}{2} +\operatorname{arccos} e^{\mu(\sqrt{2V}-1) }\Bigg).
\end{equation}


\section{Key generation rate} \label{keyrate_part}
In the previous section we have shown how HOM experiments' results could be used for upper-bounding the degree of distinguishability for both BB84 and decoy-state techniques. The final goal of this procedure is to find the secret key rate, depending on the measurement results. In this section we simulate the key generation rate for a system with BB84 protocol and two decoy states (i.e. three intensities). The asymptotic key generation rate can be lower bounded as \cite{ma2005}:

\begin{equation}
K \geq \max_{I_s,I_d} \left[   \frac{1}{2}\left( p_1^s Y_{1L}^s [1-h(e_{1U}^s)]-f(E^s) Q^s h(E^s) \right) \right].
\end{equation}
Here maximization is done among the intensities of signal ($I_s$) and decoy ($I_d$) states. The second decoy state intensity is considered to be zero. The probability of generating single photon for a signal state is $p_1^s = I_s e^{-I_s}$, according to the Poisson distribution. The lower bound for the yield of single photon for signal states is $Y_{1L}^s$, and $e_{1U}^s$ is the upper bound for single photon error rate within a signal state. $Q^s$ is the signal state gain, $f(E^s)$ is the efficiency of error correction and $h(E^s)$ is the binary Shannon entropy with the signal error rate $E^s$ as an argument.

The parameter affected by the bases imperfections is the $e_{1U}^s$, as Eve has an ability to get more information with the same overall error rate.

For simulation we use error correction coefficient $f(E^s)$ = 1.2; fiber attenuation 0.2 dB/km; Bob's device losses 3 dB; detector efficiency 25\%, optical error rate 1\% and dark count probability per gate $10^{-5}$. These values match with the ones used in previous simulations  \cite{lucamarini2015,tamaki2016}. To achieve higher key rates, we make reasonable assumption that Bob's losses are calibrated and cannot be lowered by Eve. This assumption is also quite common for QKD models \cite{maroy2010,maroy2017,huang2018}. Finally, to minimize the number of displayed parameters we substitute the same visibility values for all pairs of sources within the simulation.

To simulate the effect of bases distinguishability, we use a method developed in \cite{lucamarini2015}. The calculated imbalance is corrected taking into account the ability of Eve to use lossless channel. 

\begin{equation}
\Delta'=\frac{\Delta }{\tilde{Y}_{1L}^{s}},
\end{equation}
where $\tilde{Y}_{1L}^{s}$ is the minimum single-photon yield among two bases, obtained using the decoy-state method. Here we assume that decoy-state method doesn't have any additional vulnerabilities. Once corrected, the basis imbalance is included in the upper bound of single-photon error of signal state:

\begin{equation}
(e_{1U}^s)' = 4(1-\Delta') \Delta' (1-2e_{1U}^s)+4 (1-2 \Delta') \sqrt{\Delta'  (1-\Delta') e_{1U}^s (1-e_{1U}^s)}.
\end{equation}

With the help of the above estimations, we perform a simulation of key generation rate. Visibility values used for simulation are chosen so that they can be compared with current experimental results that we discuss in the next section.

\begin{figure} [h]
\centering
\includegraphics[width=0.75\textwidth]{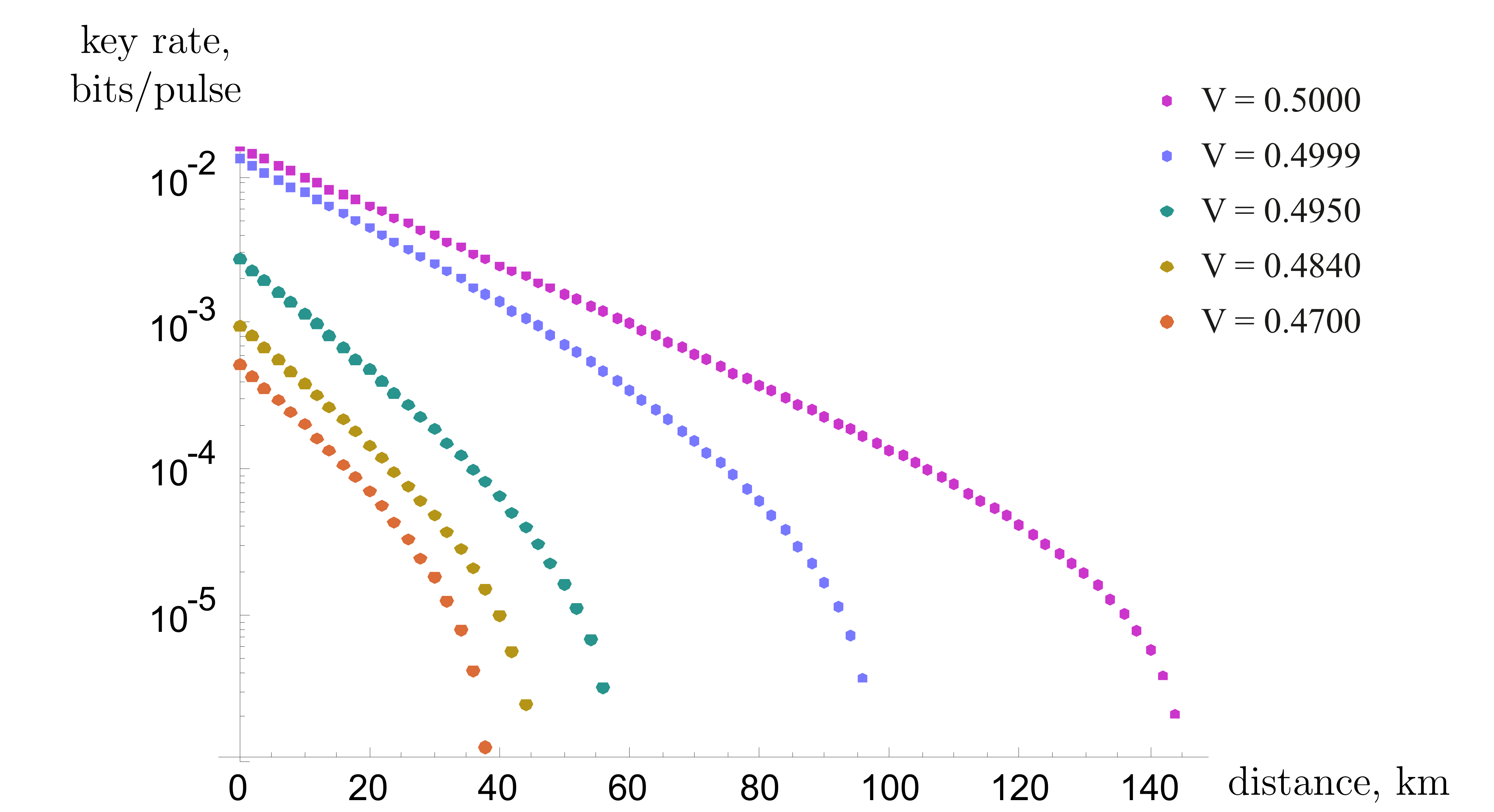}
\caption{Key generation rate depending on the communication distance for different values of HOM visibility for mode-matched states.}
\label{key_basis_flaw}
\end{figure}

Simulation results shown in the Fig. \ref{key_basis_flaw}, indicate that key generation rate is highly sensitive to the distinguishability of the bases of BB84 protocol (Fig. \ref{key_basis_flaw}). Even for pretty high values of visibility like 0.495, the key rate drops significantly. On the other hand, for the visibility as low as 0.47 key generation is still possible.

\section{Applications and discussion} \label{discussion_part}

To check applicability of the proposed method for practical setups, we take the best to date fourth-order PRWCPs interference visibilities from  various experiments carried out for the MDI-QKD \cite{da2013,rubenok2013,tang2014,comandar2016mdi,comandar2016}.

\begin{table} [h]
	\centering
\begin{tabular}{ c c } 

	 & HOM visibility\\
	\hline
	\hline
da Silva et al. \cite{da2013} & 0.478 \\
	\hline
Rubenok et al. \cite{rubenok2013} & $0.47 \pm 0.01$	\\
	\hline
Tang et al. \cite{tang2014}	& $0.475 \pm 0.010$ \\
	\hline
Comandar et al. \cite{comandar2016} (without post-selection) & $0.487 \pm 0.003$ \\
	\hline
Comandar et al. \cite{comandar2016} (with post-selection) & $0.499 \pm 0.004$ \\
	\hline	
	\hline
\end{tabular}
	\caption{Best to date HOM visibility results with PRWCPs.} \label{table_HOM}
\end{table}

According to the results of the previous section, it is clear that all listed results provide positive key rate. It should be also noted that latest results saturate theoretical limit \cite{comandar2016}. This fact allows to expect more precise experiments in the future.

As it has been mentioned \cite{comandar2016}, one of the main reasons of non-perfect interference between the pulses from independent gain-switched semiconductor laser diodes is the combination of frequency chirp and emition time jitter. While the former is a vulnerability that could be used by Eve to extract information about the states, the latter can be truly quantum as the time uncertainty includes spontaneous emission jitter. This leads to possible underestimation of the key generation rate, as quantum jitter dramatically decreases the interference visibility but does not compromise the setup.

To date, the best visibility has been achieved with the help of optical seeding, that leads to reduction of jitter from approximately 30 ps to 10 ps \cite{comandar2016}. Moreover, authors performed post-selection on the results, removing double clicks which are separated by time interval longer than FWHM of the double click peak, and improved the visibility up to $0.499 \pm 0.004$ saturating the theoretical limit. Since we try to upper-bound all possible differences between the pulses, it is not clear if this kind of post-selection can be used for distinguishability evaluation purposes. If we are sure that jitter is mainly quantum and out of Eve's control, such kind of post selection could be treated as removing the events with worst time overlap due to the quantum randomness of the jitter. In this case we make a tighter bound on the Eve's abilities. We have to be sure, however that the jitter is not caused by the classical effects, which Eve may possibly account for.

The results by Comandar et al. \cite{comandar2016} show that pulsed laser seeding is a promising technology to design a modulator-free polarization encoding device with multiple sources. Indeed, the method proposed in this paper together with such kind of design allow to construct a provably secure source, as an alternative to current ones, used in satellite QKD.


It should also be mentioned that the proposed method is limited by the single-photon detector wavelength sensitivity. If the source contains side channels in the wavelength range that is out of the detector's sensitivity, or even not electromagnetic, this kind of side channels will not be detected by the proposed approach. One of such examples is the first ever QKD prototype, where one could hear which state has been generated \cite{bennett1992first,brassard2005history} due to high-voltage power supply of Pockels cell.

\section*{Conclusion}

We introduce a  method for an integral evaluation of passive side-channel information leakage from Alice. We show that for relatively small intensities, fourth-order interference visibility linearly depends on the degree of non-orthogonality between two PRWCPs. This allows us to connect the results of Hong-Ou-Mandel experiments with bases imbalance. We include the obtained visibility values into the security model, and calculate the key generation rate. In this work we address conventional BB84 protocol with decoy states, however, the concept could be applied to other protocols as well. Current experimental data has been used to check the applicability of our method for the real-world systems. Future studies on this topic may include estimation of distinguishability between signal and decoy states. The developed method can be used both for certification of available QKD systems and for provably-secure systems' design in the future.

\section*{Acknowledgements}
The work was supported by the Russian Science Foundation  grant No. 17-71-20146 (AD) in the parts ``Side-channel information and HOM visibility'' (Sec.~\ref{basis_flaw_part}), ``Key generation rate'' (Sec.~\ref{keyrate_part}) and the Russian Science Foundation  grant No. 17-72-30036 (DS) in the part ``Hong-Ou-Mandel interference and non-orthogonality of quantum states'' (Sec.~\ref{HOM_part}).

\section*{Appendix A: Fidelity estimation}

In this section we calculate fidelity for two PRWCPs with density matrices $\hat{\rho_1}$ and $\hat{\rho_2}$, from eq. (\ref{rho}) as a function of HOM visibility.
\begin{equation} \label{rho1}
\hat{\rho_1}=\sum_{n=0}^{\infty} \rho_{nn} \ket{n}_1 \bra{n}_1 = \sum_{n=0}^{\infty} e^{-\mu^2} \frac{\mu^n}{n!} (\hat{a}_1^{\dagger})^n\ketbra{0}{0}(\hat{a}_1)^n
\end{equation},

\begin{equation}
\hat{\rho_2}= \sum_{n=0}^{\infty} \frac{\rho_{nn}}{n!}(\sqrt{\gamma}\hat{a}_1^{\dagger}+\sqrt{1-\gamma}\hat{a}_{\bot 1}^{\dagger})^n\ketbra{0}{0}(\sqrt{\gamma}\hat{a}_1+\sqrt{1-\gamma}\hat{a}_{\bot 1})^n
\end{equation}
Here we've decomposed the $a_2^{\dagger}$ operator into a sum of the $a_1^{\dagger}$ and $a_{\bot 1}^{\dagger}$. Therefore $a_{\bot 1}^{\dagger}$ represents all the modes that are orthogonal to $a_1^{\dagger}$. Parameter $\gamma$ indicates the similarity of these states.
We rewrite $\hat{\rho_2}$ as:
\begin{equation}
\hat{\rho_2}= \sum_{n=0}^{\infty} \rho_{nn} 
\Bigg( \sum_{k=0}^n \sqrt{C_n^k} (\sqrt{\gamma})^k(\sqrt{1-\gamma})^{n-k}\ket{k}_1\ket{n-k}_{\bot 1}\Bigg)\cdot
\Bigg(\sum_{q=0}^n \bra{q}_1\bra{n-q}_{\bot 1} \sqrt{C_n^q} (\sqrt{\gamma})^q(\sqrt{1-\gamma})^{n-q}\Bigg).
\end{equation}
For simplicity we replace $(1 - \gamma)$ with a new letter $\delta$.
Visualizing the density matrices $\hat{\rho_1}$ and $\hat{\rho_2}$ we have:

\begin{equation} \label{rho1vis}
\hat{\rho}_1=\mqty(\dmat{
\rho_{00},
\rho_{11}\mqty(1&0\\0&0),
\rho_{22}\mqty(1&0&0\\0&0&0\\0&0&0),
 ... } )
\end{equation}

\begin{equation} \label{rho2vis}
\setlength{\arraycolsep}{2pt}
\medmuskip = 0.5mu
\hat{\rho}_2=\mqty(\dmat{
\rho_{00},
\rho_{11}\mqty({ \gamma}&{ \sqrt{\gamma\delta}}\\
{ \sqrt{ \gamma\delta}}&{ \delta}),
\rho_{22}\mqty({ \gamma^2}&{ \sqrt{2 \gamma^3 \delta}}&{ \gamma\delta}\\
{ \sqrt{2 \gamma^3 \delta}}&{2 \gamma\delta}&{ \sqrt{2 \gamma\delta^3}}\\
{ \gamma\delta}&{ \sqrt{2 \gamma\delta^3}}&{\delta^2}),
 ... } ).
\end{equation}
According to the fidelity definition (\ref{fidelity}), we straightforwardly obtain the result:
\begin{equation}
\sqrt{F(\hat{\rho}_1, \hat{\rho}_2)} = \rho_{00} + \sum_{n=1}^{\infty} \rho_{nn} \gamma^{\frac{n}{2}} = \exp(\mu(\sqrt{\gamma}-1)).
\end{equation}
Now we address the visibility equation (\ref{visibility}) and find that $\operatorname{V}_{\hat{\rho}_1 \hat{\rho}_2} = \frac{\gamma}{2}$. So the final result is:
\begin{equation}
\sqrt{F(\hat{\rho}_1, \hat{\rho}_2)} = \exp(\mu(\sqrt{2\operatorname{V}_{\hat{\rho}_1 \hat{\rho}_2}}-1)).
\end{equation}
\section*{Appendix B: Basis imbalance estimation}
To include basis flaw into security model one needs to calculate fidelity between two bases ($F(\hat{\rho}_x, \hat{\rho}_z)$). Density matrices of X and Z bases could be expressed as follows:

\begin{equation}
\hat{\rho}_x = 
\frac{1}{2}\mqty(\dmat{1 & 0 \\ 0 & 0}) \otimes \hat{\rho}_{x, 0}^{\lambda} +
\frac{1}{2}\mqty(\dmat{0 & 0 \\ 0 & 1}) \otimes \hat{\rho}_{x, 1}^{\lambda},
\end{equation}

\begin{equation}
\hat{\rho}_z = 
\frac{1}{2}\mqty(\dmat{ 1/2 &  1/2 \\ 1/2 &  1/2}) \otimes \hat{\rho}_{z, 0}^{\lambda} +
\frac{1}{2}\mqty(\dmat{ 1/2 &  - 1/2 \\ - 1/2 &  1/2}) \otimes \hat{\rho}_{z, 1}^{\lambda}.
\end{equation}

Here, $\hat{\rho}_{i, j}^{\lambda}$ represents density matrices of non-operational degrees of freedom.

\begin{figure} [h]
	\centering
	\includegraphics[width=0.6\textwidth]{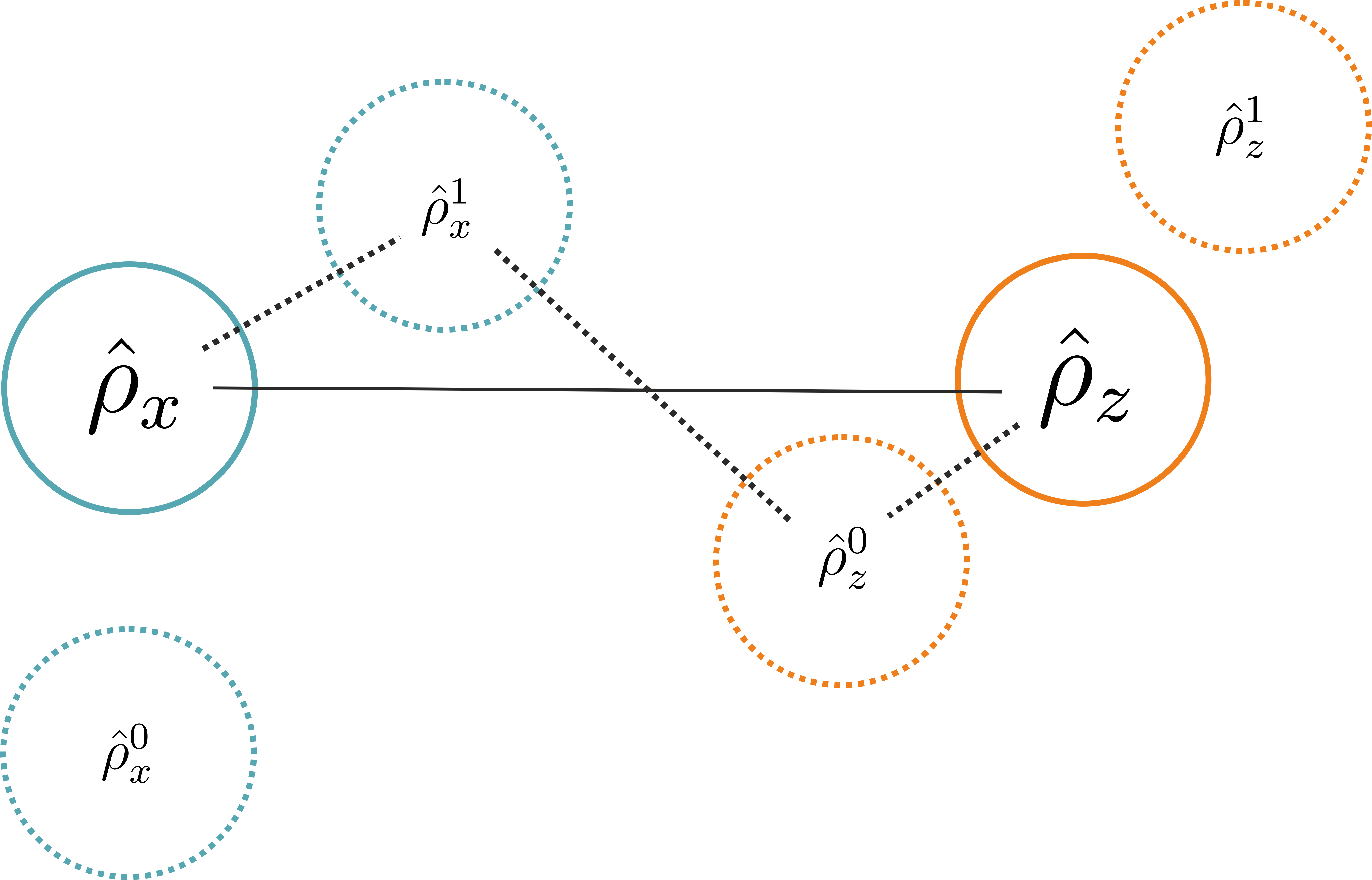}
	\caption{Scheme of bases fidelity estimation. A pair of auxiliary matrices for each basis allow to bound the distance using two triangle inequalities. The distance between two closest auxiliary matrices is used.}
	\label{fidelity_scheme}
\end{figure}

We calculate the fidelity between the bases in two steps. First, we introduce a set of auxiliary density matrices, calculate the fidelity between the initial and auxiliary states and then, between auxiliary matrices for different bases. After that, we use the distance and apply triangle inequities, that allow us to bound the fidelity between the initial states. The idea of the estimation is illustrated om the figure \ref{fidelity_scheme}.  

 We define auxiliary density matrices as bases, for which non-operational degrees of freedom are same for both bits:

\begin{equation} \label{ideal_dens}
\hat{\rho}^j_i = 
\frac{1}{2}\mqty(\dmat{1 & 0 \\ 0 & 0}) \otimes \hat{\rho}_{i, j}^{\lambda} +
\frac{1}{2}\mqty(\dmat{0 & 0 \\ 0 & 1}) \otimes \hat{\rho}_{i, j}^{\lambda}.
\end{equation}

We calculate the fidelity between $\hat{\rho}_x$ and an auxiliary matrix, where for both bits are the same in non-operational degrees of freedom. Without loss of generality, we start with a case, when both bits have a density matrix  $\hat{\rho}_{x, 0}^{\lambda}$ for non-operational parameters: $\hat{\rho}_x^{0} = \frac{1}{2} \mqty(\dmat{1 & 0 \\ 0 & 1}) \otimes \hat{\rho}_{x, 0}^{\lambda}$.

\begin{equation}
F(\hat{\rho}_x, \hat{\rho}_x^{0}) = F \Bigg(
\frac{1}{2}\mqty(\dmat{\hat{\rho}_{x, 0}^{\lambda} & 0 \\ 0 & \hat{\rho}_{x, 1}^{\lambda}}) ,
\frac{1}{2}\mqty(\dmat{\hat{\rho}_{x, 0}^{\lambda} & 0 \\ 0 & \hat{\rho}_{x, 0}^{\lambda}}) 
\Bigg).
\end{equation}

As a result, using fidelity definition, we obtain:

\begin{equation} \label{fidX}
\sqrt{F(\hat{\rho}_x, \hat{\rho}_x^{0})} = 
\frac{1+\sqrt{F(\hat{\rho}_{x, 0}^{\lambda}, \hat{\rho}_{x, 1}^{\lambda})}}{2} =
\sqrt{F(\hat{\rho}_x, \hat{\rho}_x^{1})}.
\end{equation}

The fidelities between the initial $\hat{\rho}_x$ and both auxiliary states $\hat{\rho}_{x}^{0}$ and $\hat{\rho}_{x}^{1}$ are the same. A similar result can be obtained for Z-basis as Hadamard transformation can be applied to the operational degree of freedom for $\hat{\rho}_z$, so the resulting fidelity is the same:

\begin{equation} \label{fidZ}
\sqrt{F(\hat{\rho}_z, \hat{\rho}_z^{0})} = 
\frac{1+\sqrt{F(\hat{\rho}_{z, 0}^{\lambda}, \hat{\rho}_{z, 1}^{\lambda})}}{2} =
\sqrt{F(\hat{\rho}_z, \hat{\rho}_z^{1})}.
\end{equation}

Now, to apply triangle inequality, we use Bures angle \cite{ma2009fidelity}, which constitutes metric in Hilbert space:
\begin{equation} 
\operatorname{A}(\hat{\rho}_1, \hat{\rho}_2)= \arccos \sqrt{F(\hat{\rho}_1, \hat{\rho}_2)}.
\end{equation}

As soon as we found the fidelity values between the initial states and the auxiliary ones, we need look for a pair of auxiliary matrices from both bases, that have maximum fidelity (minimum distance). 

As a result of two triangle inequalities we obtain a result:
\begin{equation} 
\arccos \sqrt{\operatorname{F}(\hat{\rho}_x, \hat{\rho}_z)} \leq \operatorname{arccos}\max_{\scriptscriptstyle i,j \in {0, 1}} \sqrt{F(\hat{\rho}_{x, i}^{\lambda}, \hat{\rho}_{z, j}^{\lambda})} + \sum_{ \scriptscriptstyle b \in \left\lbrace x, z \right\rbrace  }  \operatorname{arccos}\frac{1+\sqrt{F(\hat{\rho}_{b, 0}^{\lambda}, \hat{\rho}_{b, 1}^{\lambda})}}{2}.
\end{equation}

The final result is expressed as follows:

\begin{equation} \label{finalDelta}
\Delta \leq \frac{1}{2} - \frac{1}{2}\cos \Bigg(\operatorname{arccos}\max_{\scriptscriptstyle i,j \in {0, 1}} \sqrt{F(\hat{\rho}_{x, i}^{\lambda}, \hat{\rho}_{z, j}^{\lambda})} + \sum_{ \scriptscriptstyle b \in \left\lbrace x, z \right\rbrace  }  \operatorname{arccos}\frac{1+\sqrt{F(\hat{\rho}_{b, 0}^{\lambda}, \hat{\rho}_{b, 1}^{\lambda})}}{2}\Bigg).
\end{equation}

In the special case when all fidelities equal the same value F, equation (\ref{finalDelta}) simplifies:
\begin{equation} \label{simpleFinalDelta}
\Delta \leq \frac{1}{2}  - \frac{1}{2}\cos\Bigg(2 \operatorname{arccos}\frac{1+\sqrt{F}}{2} +\operatorname{arccos}\sqrt{F}\Bigg).
\end{equation}

\bibliography{ref}
\bibliographystyle{unsrt}

\end{document}